\RequirePackage{fix-cm}
\documentclass[smallextended]{svjour3}       % onecolumn (second format)
\smartqed  % flush right qed marks, e.g. at end of proof
\usepackage{graphicx}
\usepackage{amsmath}
\usepackage{amsfonts}
\usepackage{epstopdf}
\usepackage{amsfonts}
\usepackage{color}
\usepackage{mathtools}
\usepackage{booktabs} % for much better looking tables
\usepackage{array} % for better arrays (eg matrices) in maths
\usepackage{verbatim} % adds environment for commenting out blocks of text & for better %verbatim
\usepackage{subfig}
\usepackage{amsmath}

\usepackage{natbib}
\usepackage{etex}
\usepackage[ruled,vlined,linesnumbered]{algorithm2e}
\newcommand{\umu}{v_m}

\newcommand{\rate}{{\left[1+ \xi \umu \right]}}

\newcommand{\sumr}{\sum_{j=1}^{r}}
\newcommand{\sumR}{\sum_{j=1}^{R}}

\newcommand{\xmu}{z_{j,m}}

\newcommand{\xrate}{\left[1+ \xi \xmu\right]}

\newcommand{\zrate}{\left[1+ \xi z\right]}

 \usepackage{mathptmx}      % use Times fonts if available on your TeX system
\usepackage{latexsym}
\begin{document}

\title{A Poisson process reparameterisation for Bayesian inference for extremes%\thanks{Grants or other notes
}


\author{Paul Sharkey         \and
        Jonathan A. Tawn %etc.
}


\institute{P. Sharkey, J.A. Tawn  \at
              STOR-i Centre for Doctoral Training,
              Department of Mathematics and Statistics,
              Lancaster University, Lancaster, LA1 4YF, United Kingdom. \\
              \email{p.sharkey1@lancs.ac.uk}           %  \\
         %  \and
         %  S. Author \at
          %    second address
}

\date{Received: date / Accepted: date}

\maketitle

\begin{abstract}
A common approach to modelling extreme values is to consider the excesses above a high threshold as realisations of a non-homogeneous Poisson process. While this method offers the advantage of modelling using threshold-invariant extreme value parameters, the dependence between these parameters makes estimation more difficult. We present a novel approach for Bayesian estimation of the Poisson process model parameters by reparameterising in terms of a tuning parameter $m$. This paper presents a method for choosing the optimal value of $m$ that near-orthogonalises the parameters, which is achieved by minimising the correlation between the asymptotic posterior distribution of the parameters. This choice of $m$ ensures more rapid convergence and efficient sampling from the joint posterior distribution using Markov Chain Monte Carlo methods. Samples from the parameterisation of interest are then obtained by a simple transform. Results are presented in the cases of identically and non-identically distributed models for extreme rainfall in Cumbria, UK.

%Insert your abstract here. Include keywords, PACS and mathematical
%subject classification numbers as needed.
\keywords{Poisson processes \and extreme value theory \and Bayesian inference \and reparameterisation \and covariate modelling}
\end{abstract}

\section{A Poisson Process model for Extremes}
\label{sec:pp}
The aim of extreme value analysis is to model rare occurrences of an observed process to extrapolate to give estimates of the probabilities of unobserved levels. In this way, one can make predictions of future extreme behaviour by
estimating the behaviour of the process using an asymptotically justified limit model. Let $X_1, X_2, \hdots, X_n$ be a series of independent and identically distributed (iid) random variables with common distribution function $F$.
Defining $M_n = \max\{X_1,X_2,\hdots,X_n\}$, if there exists sequences of normalising constants $a_n > 0$ and $b_n$ such that:
\begin{equation}
 \Pr\left\{\frac{M_n - b_n}{a_n} \leq x \right\} \rightarrow G(x) \hspace{15pt} \mbox{             } \text{as} \mbox{          } n \rightarrow \infty,
 \label{eq:asymp}
\end{equation}
where $G$ is non-degenerate,
then $G$ follows a generalised extreme value (GEV) distribution, with distribution function
\begin{equation}
 G(x) = \exp \left\{- { \left[1+\xi \left( \frac{x-\mu}{\sigma} \right) \right] }_{+}^{-1/\xi} \right\},
 \label{eq:gev}
\end{equation}
where $x_{+} = \max(x,0)$, $\sigma >0$ and $\mu, \xi \in \mathbb{R}$. Here, $\mu, \sigma$ and $\xi$ are location, scale and shape parameters respectively. \\

Using a series of block maxima from $X_1, \hdots, X_n$, typically with blocks corresponding to years, the standard inference approach to give estimates of $(\mu,\sigma,\xi)$ is the maximum likelihood technique, which requires numerical optimisation methods. In these problems, particularly when covariates are involved, such methods may converge to local optima, with the consequence that parameter estimates are largely influenced by the choice of starting values.  
The standard asymptotic properties of the maximum likelihood estimators are subject to certain regularity conditions outlined in \citet{smith1985maximum}, but can give a poor representation of true uncertainty. In addition, flat likelihood surfaces can cause identifiability issues \citep{smith1987parameter}. 
%In a likelihood framework, numerical maximisation schemes tend
%to converge to local optima, with the consequence that parameter estimates are largely influenced by choice of starting values. \citet{smith1987parameter} demonstrated that in the likelihood framework, identifiability issues may arise from flat likelihood surfaces even under orthogonal parameterisations. In such
%circumstances, asymptotic results for maximum likelihood estimators give a poor representation of true uncertainty. 
For these reasons, we choose to work in a Bayesian setting.
Bayesian approaches have been used to make inferences about $\boldsymbol{\theta} = (\mu,\sigma,\xi)$ using standard Markov Chain Monte Carlo (MCMC) techniques. They have the advantage of being able to incorporate prior information when little is known about the extremes of interest, while also better accounting for parameter uncertainty when estimating functions of $\boldsymbol{\theta}$, such as return levels \citep{coles1996bayesian}. For a recent review, see \citet{stephenson2016bayesian}.
 \\

An approach to inference that is considered to be more efficient than using block maxima is to consider a model for threshold excesses, which is superior in the sense that it reduces uncertainty due to utilising more extreme data \citep{smithreport}. Given a high threshold $u$, the conditional distribution of excesses above $u$ can be approximated by a generalised Pareto (GP) distribution \citep{pickands1975statistical} such that
\[ \Pr(X-u > x | X > u) = {\left(1+\frac{\xi x}{\psi_{u}}\right)}^{-1/\xi}_{+}, \mbox{                                } x > 0,\]
where $\psi_u > 0$ and $\xi \in \mathbb{R}$ denote the scale and shape parameters respectively, with $\psi_u$ dependent on the threshold $u$, while $\xi$ is identical to the shape parameter of the GEV distribution. This model conditions on an exceedance, but a third parameter $\lambda_u$, denoting the rate of exceedance of $X$ above the threshold $u$, must also be estimated. \\

Both of these extreme value approaches are special cases of a unifying limiting Poisson process characterisation of extremes \citep{smith1989extreme,coles2001introduction}. 
Let $P_n$ be a sequence of point processes such that
\[ P_n = \left\{ \left(\frac{i}{n+1} , \frac{X_i - b_{n}}{a_{n}} \right) : i = 1, \hdots, n \right\}, \]
where $a_n>0$ and $b_n$ are the normalising constants in limit (\ref{eq:asymp}). The limit process is non-degenerate since the limit distribution of $(M_n-b_n)/a_n$ is non-degenerate.
Small points are normalised to the same value $b_L = \lim_{n \to \infty} (x_L - b_n)/a_n$, where $x_L$ is the lower endpoint of the distribution $F$. Large points are retained in the limit process.
It follows that $P_n$ converges to a non-homogeneous Poisson process $P$ on regions of the form $A_y = (0,1) \times [y, \infty)$, for $y>b_L$. The limit process $P$ has an intensity measure on $A_y$ given by
\begin{equation}
\Lambda(A_y) = {{\left[1+ \xi\left(\frac{y-\mu}{\sigma}\right)\right]}}_{+}^{-1/\xi}.
\label{eq:intens}
\end{equation}
It is typical to assume that the limit process is a reasonable approximation to the behaviour of $P_n$, without normalisation of the $\{X_i\}$, on $A_u = (0,1) \times [u, \infty)$, where $u$ is a sufficiently high threshold and $a_n$, $b_n$ are absorbed into the location and scale parameters of the intensity (\ref{eq:intens}). It is often convenient to rescale the intensity by a factor $m$, where $m > 0$ is free, so that the $n$ observations consist of $m$ blocks of size $n/m$ with the maximum $M_m$ of each block following a GEV$(\mu_m, \sigma_m, \xi)$ distribution, with $\xi$ invariant to the choice of $m$. The Poisson process likelihood can be expressed as
\begin{equation}
 L(\boldsymbol{\theta}_m) = \exp\left\{-m {{\left[1+ \xi \left(\frac{u-\mu_m}{\sigma_m}\right)\right]}}_{+}^{-1/\xi} \right\} \prod_{j=1}^{r} \frac{1}{\sigma_m} {{\left[1+ \xi \left(\frac{x_j-\mu_m}{\sigma_m}\right)\right]}}_{+}^{-1/\xi -1},
 \label{eq:like}
\end{equation}
where $\boldsymbol{\theta}_m = (\mu_{m},\sigma_{m},\xi)$ denotes the rescaled parameters, $r$ denotes the number of excesses above the threshold $u$ and $x_j > u, j = 1, \hdots, r$, denote the exceedances. 
It is possible to move between parameterisations associated with different numbers of blocks. If for $k$ blocks the block maximum is denoted by $M_k$ and follows a GEV distribution with the parameters $\boldsymbol{\theta}_k = (\mu_{k},\sigma_{k},\xi)$, then for all $x$
\[ \Pr(M_k < x) = \Pr{(M_m < x)}^{k/m}. \]
As $M_k$ is GEV$(\mu_{k},\sigma_{k},\xi)$ and $M_m$ is GEV$(\mu_{m},\sigma_{m},\xi)$ it follows that 
%\begin{equation}
% \Lambda(A) = k{\left[1+ \xi_m \left(\frac{u-\mu_m}{\sigma_m}\right) \right]}^{-1/\xi_m},
% \label{eq:mintens}
%\end{equation}
%which holds for all $k$, then $\boldsymbol{\theta}_k$ can be expressed in terms of $\boldsymbol{\theta}_m$ through the expressions
\begin{eqnarray}
 \mu_k & = & \mu_m - \frac{\sigma_m}{\xi} \left(1-{\left(\frac{k}{m}\right)}^{-\xi} \right) \nonumber \\
 \sigma_k & = & \sigma_m {\left(\frac{k}{m}\right)}^{-\xi}  \label{eq:trans}.
\end{eqnarray}
In this paper, we present a method to improve inference for $\boldsymbol{\theta}_k$, the parameterisation of interest. For an `optimal' choice
of $m$ we first undertake inference for $\boldsymbol{\theta}_m$ before transforming our results to give inference for $\boldsymbol{\theta}_k$ using the mapping in expression (\ref{eq:trans}). \\
%Section~\ref{sec:bayes} describes how the mapping in (\ref{eq:trans}) can be used to improve estimation of $\boldsymbol{\theta}_k$ for a given $k$ through the choice of an `optimal' $m$. \\

In many practical problems, $k$ is taken to be $n_y$, the number of years of observation, so that the annual maximum has a GEV distribution with parameters $\boldsymbol{\theta}_{n_y}=(\mu_{n_y},\sigma_{n_y},\xi)$. %, which we denote by $(\mu,\sigma,\xi)$.
Although inference is for the annual maximum distribution parameters $\boldsymbol{\theta}_{n_y}$, the Poisson process model makes use of all data that are extreme, so inferences are more precise than estimates based on a direct fit of the GEV distribution to the annual maximum data as noted above. \\

To help see how the choice of $m$ affects inference, consider the case when $m=r$, the number of excesses above the threshold $u$. If a likelihood inference was being used with this choice of $m$, the maximum likelihood estimators $(\hat{\mu}_r,\hat{\sigma}_r,\hat{\xi}) = (u, \hat{\psi}_u, \hat{\xi})$, 
% The shape parameter of the Poisson process model is equal to that of the GP, while the scale parameter is related through the identity
% \begin{equation} 
% \psi_u = \sigma_m + \xi (u-\mu_m),
% \label{eq:sigid}
% \end{equation}
% which is invariant to choice of $m$. Let $R_u$ denote the number of excesses above the threshold $u$. Then $R_u$ is a Poisson random variable with expectation given by the intensity measure of the point process, such that
% \begin{equation}
%  \mathbb{E}(R_u) = m {{\left[1+ \xi \left(\frac{u-\mu_m}{\sigma_m}\right)\right]}}_{+}^{-1/\xi}.
% \end{equation}
% By setting $m = r$, we get $\hat{\mu}_r = u$. 
see Appendix~\ref{App:gp} for more details. 
%It follows from identity (\ref{eq:sigid}) that the estimated scale parameters of the two models are equal, i.e. $\hat{\psi}_u = \hat{\sigma}_r$. 
Therefore, Bayesian inference for the parameterisation of the Poisson process model when $m=r$ is equivalent to Bayesian inference for the GP model. \\

Although inference for the Poisson process and GP models is essentially the same approach when $m=r$, they differ in parameterisation, and hence inference, when $m \neq r$.
The GP model is advantageous in that $\lambda_u$ is globally orthogonal to $\psi_u$ and $\xi$. \citet{chavez2005generalized} achieved local orthogonalisation of the GP model at the maximum likelihood estimates by reparameterising the scale parameter as
$\nu_u = \psi_u (1+\xi)$. This ensures all the GP tail model parameters are orthogonal locally at the likelihood mode. However, the scale parameter is still dependent on the choice of threshold.
Unlike the GP, the parameters of the Poisson process model are invariant to choice of threshold, which makes it more suitable for covariate modelling
and hence suggests that it may be the better parameterisation to use. In contrast, it has been found that the parameters are highly dependent, making estimation more difficult. \\

As we are working in the Bayesian framework, strongly dependent parameters lead to poor mixing in our MCMC procedure \citep{hills1992parameterization}. A common way of overcoming this is to explore the parameter space using
a dependent proposal random walk Metropolis-Hastings algorithm, though this requires a knowledge of the parameter dependence structure \emph{a priori}. Even in this case, the dependence structure potentially varies in different regions of the parameter space,
which may require different parameterisations of the proposal to be applied. The alternative approach is to consider a reparameterisation to give orthogonal parameters. However, \citet{cox1987parameter} show that global orthogonalisation cannot be achieved in general. \\

This paper illustrates an approach to improving Bayesian inference and efficiency for the Poisson process model. Our method exploits the scaling factor $m$ as a means of creating
a near-orthogonal representation of the parameter space. 
%Parameter orthogonality can be achieved by
%reparameterising the model through the choice of $m$ so that the Fisher information matrix is diagonal.  \\
While it is not possible in our case to find a value of $m$ that diagonalises the Fisher information matrix, we focus on minimising the off-diagonal components of the covariance matrix. We present a method for
choosing the `best' value of $m$ such that near-orthogonality of the model parameters is achieved, and thus improves the convergence of MCMC and sampling from the joint posterior distribution. Our focus is on Bayesian inference but the reparameterisations we find can be used to improve likelihood inference as well, simply by ignoring the prior term. \\

The structure of the paper is as follows. Section~\ref{sec:bayes} examines the idea of reparameterising in terms of the scaling factor $m$ and how this can
be implemented in a Bayesian framework. Section~\ref{sec:choose} discusses the choice of $m$ to optimise the sampling from the joint posterior distribution in the case where $X_1, \hdots, X_n$ are iid. Section~\ref{sec:nonstat} explores
this choice when allowing for non-identically distributed variables through covariates in the model parameters. Section~\ref{sec:cumbria} describes an application of our methodology to extreme rainfall in Cumbria, UK, which experienced major flooding events in November 2009 and December 2015.

\section{Bayesian Inference}
\label{sec:bayes}
Bayesian estimation of the Poisson process model parameters involves the specification of a prior distribution $\pi(\boldsymbol{\theta}_m)$. Then using Bayes Theorem, the posterior distribution of $\boldsymbol{\theta}_m$ can be expressed as 
\[ \pi(\boldsymbol{\theta}_m|\mathbf{x}) \propto \pi(\boldsymbol{\theta}_m) L(\boldsymbol{\theta}_m), \]
where $L(\boldsymbol{\theta}_m)$ is the likelihood as defined in (\ref{eq:like}) and $\mathbf{x}$ denotes the excesses of the threshold $u$.
We sample from the posterior distribution using a random walk Metropolis-Hastings scheme. Proposal values of each parameter are drawn sequentially from a univariate Normal distribution and accepted with a probability defined as the posterior ratio of the proposed state relative to the current state of the Markov chain. In all cases throughout the paper, each individual parameter chain is tuned to give the acceptance rate in the range of $20\%-25\%$ to satisfy the optimality criterion of \citet{roberts2001optimal}. 
For illustration purposes, results in Sections~\ref{sec:bayes} and~\ref{sec:choose} are from the analysis of simulated iid data.
A total of $300$ exceedances above a threshold $u=30$ are simulated from a Poisson process model with $\boldsymbol{\theta}_1 = (80,15,0.05)$.
\begin{figure}[h!]
\centering
 \includegraphics[width=6cm,angle=270]{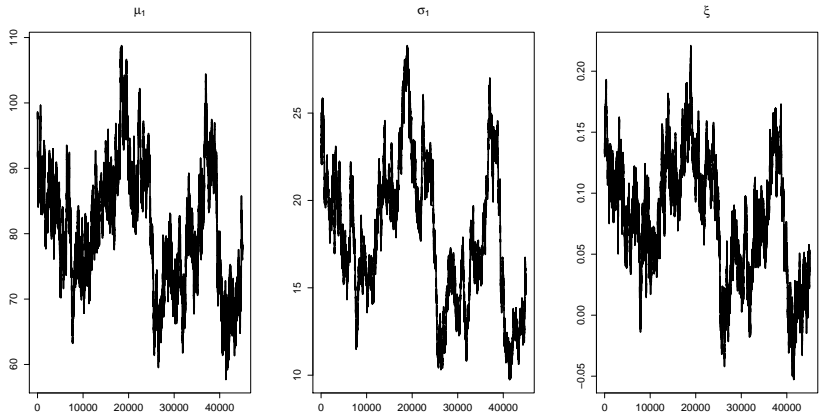}
 \caption{Random-walk Metropolis chains run for each component of $\boldsymbol{\theta}_1$.}
 \label{fig:mixing1}
\end{figure}
Figure \ref{fig:mixing1} shows individual parameter chains for $\boldsymbol{\theta}_k$ from a random walk Metropolis scheme run for $50,000$ iterations with a burn-in of $5,000$ removed, where $k=1$ and a chosen $m=1$. This figure shows the clear poor mixing of each component of $\boldsymbol{\theta}_1$, indicating non-convergence and strong dependence
in the posterior sampling. \\

\begin{figure}[h!]
\centering
 \includegraphics[width=6cm,angle=270]{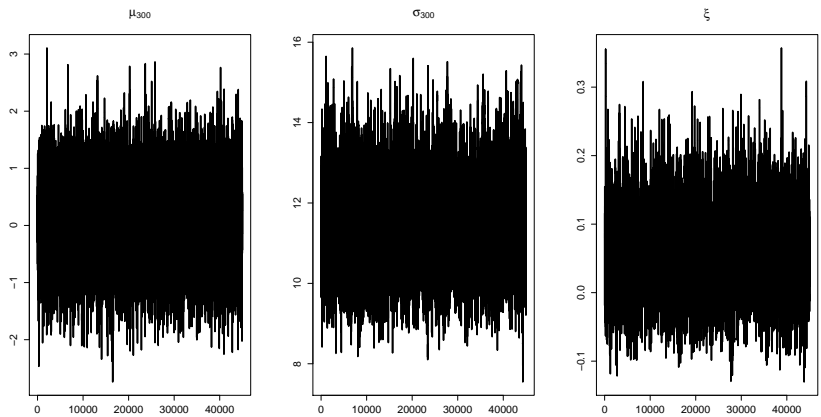}
 \caption{Random-walk Metropolis chains run for parameters $\boldsymbol{\theta}_{r}$, where $r=300$ is the number of exceedances in the simulated data.}
 \label{fig:mixing2}
\end{figure}

We explore how reparameterising the model in terms of $m$ can improve sampling performance. For a general prior on the parameterisation of interest $\boldsymbol{\theta}_k$, denoted by $\pi(\boldsymbol{\theta}_k)$, Appendix B derives that the prior on the transformed parameter space $\boldsymbol{\theta}_m$ is
\begin{equation}
 \pi(\boldsymbol{\theta}_m) = \left(\frac{m}{k}\right)^{-\xi} \pi(\boldsymbol{\theta}_k).
 \label{eq:prior}
\end{equation}
In this example, independent Uniform priors are placed on $\mu_1$, $\log \sigma_1$ and $\xi$, which gives
\begin{equation}
 \pi(\boldsymbol{\theta}_1) \propto \frac{1}{\sigma_1}; \mbox{          } \hspace{10pt} \mu_1 \in \mathbb{R}, \sigma_1 >0, \xi \in \mathbb{R}.
 \label{eq:prior}
\end{equation}
This choice of prior results in a proper posterior distribution, provided there are at least 4 threshold excesses \citep{northropproper}.
By finding a value of $m$ that near-orthogonalises the parameters of the posterior distribution $\pi(\boldsymbol{\theta}_m | \boldsymbol{x})$, we can run an efficient MCMC scheme on $\boldsymbol{\theta}_m$ before transforming the samples to $\boldsymbol{\theta}_k$.
It is noted in \citet{wadsworth2010accounting} that setting $m$ to be the number of exceedances above the threshold, i.e. $m=r$, improves the mixing properties of the chain, as is illustrated in Figure \ref{fig:mixing2}. This is approximately equivalent to inference using a GP model, as discussed in Section~\ref{sec:pp}.
\\

\begin{figure}[h!]
\centering
 \includegraphics[width=8cm,angle=270]{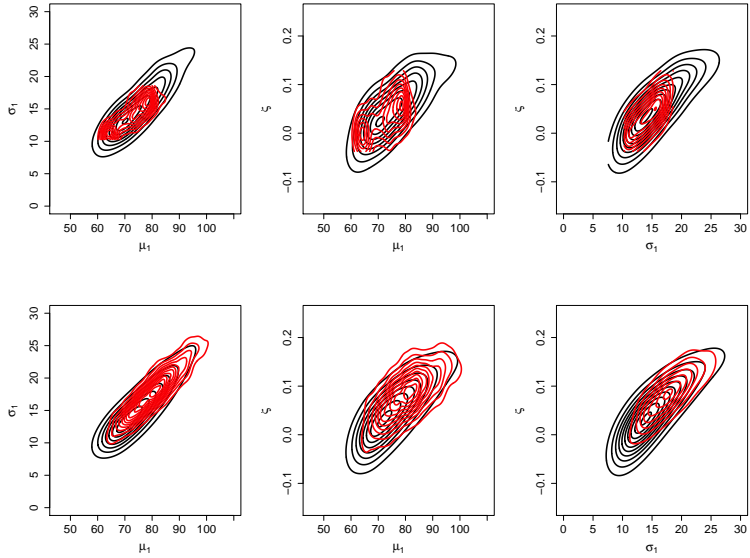}
 \caption{Contour plots of the estimated joint posterior of $\boldsymbol{\theta}_1$ for 4,000 iterations (top) and 45,000 iterations (bottom) created from the transformed samples drawn from the MCMC procedure for  $\boldsymbol{\theta}_m$ (in black) and samples of $\boldsymbol{\theta}_1$ drawn directly (in red).}
 \label{fig:corplot}
\end{figure}

Given this choice of $m$, the MCMC scheme is run for $\boldsymbol{\theta}_m$ before transforming to estimate the posterior of $\boldsymbol{\theta}_1$ using the mapping in (\ref{eq:trans}), where $k=1$ in this case. Figure \ref{fig:corplot} shows contour plots of estimated joint posterior densities of $\boldsymbol{\theta}_1$ based on 5,000 and 50,000 run lengths, with burn-in periods of 1,000 and 5,000 respectively. It compares the samples from directly estimating the posterior of $\boldsymbol{\theta}_1$ with that from transforming from the MCMC samples of the posterior of $\boldsymbol{\theta}_m$ to give a posterior sample for $\boldsymbol{\theta}_1$. Figure \ref{fig:corplot} indicates that $\boldsymbol{\theta}_1$ are highly correlated, with the result that we only sample from a small proportion of the parameter
space when exploring using independent random walks for each parameter. This explains the poor mixing if we were to run the MCMC without a transformation. In particular, very different estimates of the joint posterior are achieved for the 5,000 and 50,000 run lengths. Even with 50,000 iterations the estimated density contours are very rough, indicating considerable Monte Carlo noise as a result of poor mixing. In contrast, it is clear that, after back-transforming to $\boldsymbol{\theta}_1$, the reparameterisation enables a more thorough exploration of the parameter space, with almost identical estimated joint density contours based on both 5,000 and 50,000 iterations. This shows a very rapid mixing of the associated MCMC. In fact, we found that the reparameterisation yielded smoother density contours for $5,000$ iterations than for 5 million iterations without the transformation. However, while this transformation is a useful tool in enabling an efficient
Bayesian inference procedure, further investigation is necessary in the choice of $m$ to achieve near-orthogonality of the parameter space and thus maximising the efficiency of the MCMC procedure.

\section{Choosing $m$ optimally}
\label{sec:choose}
As illustrated in Section~\ref{sec:bayes}, the choice of $m$ in the Poisson process likelihood can improve the performance of the MCMC required to estimate the posterior density of model parameters $\boldsymbol{\theta}_k$. We desire a value of $m$ such that
near-orthogonality of $\boldsymbol{\theta}_m$ is achieved, before using the expressions in (\ref{eq:trans}) to transform to the parameterisation of interest, e.g. $\boldsymbol{\theta}_1$ or $\boldsymbol{\theta}_{n_y}$. As a measure of dependence, we use the asymptotic expected correlation matrix of the posterior distribution of $\boldsymbol{\theta}_m | \boldsymbol{x}$. In particular, we explore how the off-diagonal components of the matrix, that is, the correlation between parameters, changes
with $m$. The covariance matrix associated with $\boldsymbol{\theta}_m | \boldsymbol{x}$ can be derived analytically by inverting the Fisher information matrix of the Poisson process log-likelihood (see Appendix~\ref{App:Stat}). The correlation matrix is then obtained by normalising so that the matrix has a unit diagonal. \\

Other choices for the measure of the dependence of the posterior could have been used, such as the inverse of the Hessian matrix (or the expected Hessian matrix) of the log-posterior, evaluated at the posterior mode. For inference problems with strong information from the data relative to the prior there will be limited differences in the approach and similar values for the optimal $m$ will be found. In contrast, if the prior is strongly informative and the number of threshold exceedances is small then the choice of $m$ from using our approach could be far from optimal. Also the use of the observed, rather than expected, Hessian may better represent the actual posterior distribution of $\boldsymbol{\theta}_m$ and deliver a choice of $m$ that better achieves orthogonalisation, see \citet{efron1978assessing} and \citet{tawn1987parameter} respectively. \\

We prefer our choice of measure of dependence as for iid problems it gives closed form results for $m$ which can be used without the computational work required for other approaches, and this gives valuable insight into the choice of $m$ to guide future implementation without the need for detailed computation of an optimal $m$. Furthermore, informative priors rarely arise in extreme value problems, and so information in the data typically dominates information in the prior, particularly around the posterior mode. It should be pointed out however, that the prior is used in the MCMC so there is no loss of prior information in our approach. Also standard MCMC diagnostics should be used even after the selection of an optimal $m$, so if the asymptotic posterior correlations differ much from the posterior correlations, making our choice of $m$ poor, this will be obvious and a more complete but computationally burdensome analysis can be conducted using the methods described above. \\

In this section, we use the data introduced in Section~\ref{sec:bayes}. For all integers $m \in [1,500]$, maximum posterior mode estimates $\hat{\boldsymbol{\theta}}_m$ are computed and pairwise asymptotic posterior correlations calculated
by substituting $\hat{\boldsymbol{\theta}}_m$ into the expressions for the Fisher information matrix, in Appendix~\ref{App:Stat}, and taking the inverse. Figure~\ref{fig:covmplot} shows how parameter
correlations change with the choice of $m$, illustrating that the asymptotic posterior distributions of $\mu_m$ and $\xi$ are orthogonal when $m=r$, the number of excesses above a threshold, which explains the findings of \citet{wadsworth2010accounting}. \\
\begin{figure}[h!]
\centering
\includegraphics[width=6cm,angle=270]{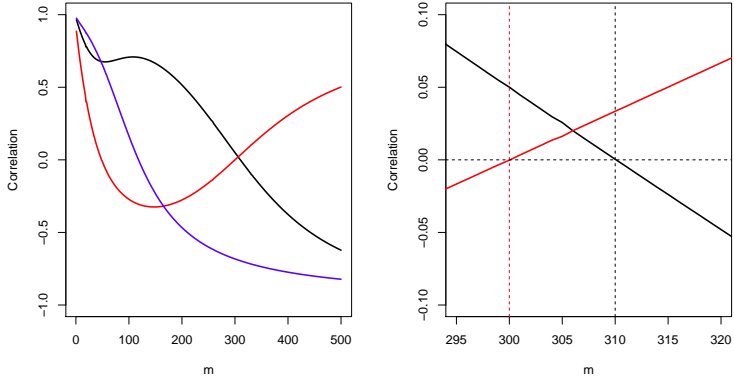}
\caption{Left: Estimated parameter correlations changing with $m$: $\rho_{\mu_m,\sigma_m}$ (black), $\rho_{\mu_m,\xi}$ (red), $\rho_{\sigma_m,\xi}$ (blue). Right: Expanded region of the graph showing $\rho_{\mu_{m},\xi} = 0$ for $m$ close to $r$ where $r=300$ is the number of excesses above the threshold, while
$\rho_{\mu_m,\sigma_m} = 0$ when $m \approx 310$.}
\label{fig:covmplot}
\end{figure}

It is proposed that MCMC mixing can be further improved by minimising the overall correlation in the asymptotic posterior distribution of $\boldsymbol{\theta}_m$. Therefore, we would like to find the value of $m$ such that $\rho({\boldsymbol{\theta}_m})$ is minimised, where $\rho({\boldsymbol{\theta}_m})$ is defined as
\begin{equation}
 \rho({\boldsymbol{\theta}_m}) = |\rho_{\mu_m,\sigma_m}| + |\rho_{\mu_m,\xi}| + |\rho_{\sigma_m,\xi}|,
 \label{eq:totcor}
\end{equation}
where $\rho_{\mu_m,\sigma_m}$ denotes the asymptotic posterior correlation between $\mu_m$ and $\sigma_m$ for example.
\begin{figure}[h!]
\centering
\includegraphics[width=8cm,angle=270]{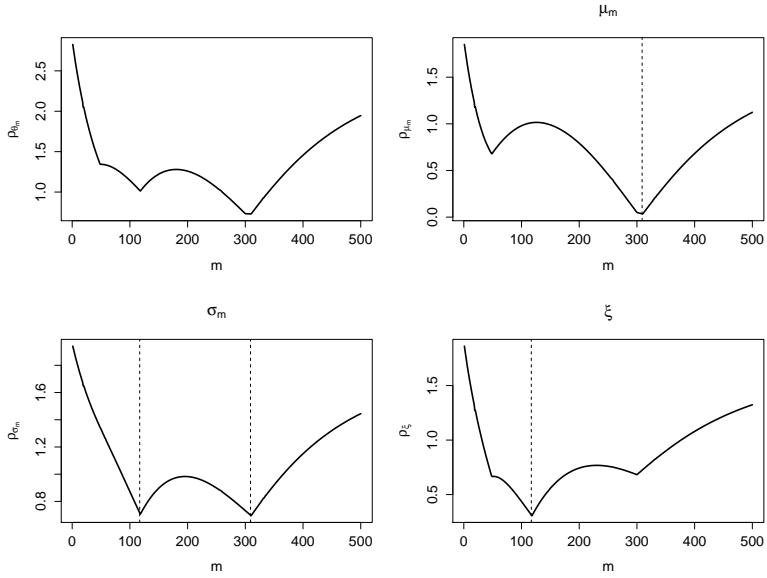}
\caption{How $\rho({\boldsymbol{\theta}_m})$ changes with $m$ (top left) and how correlations in each individual estimated parameter, as measured by $\rho_{\mu_m}, \rho_{\sigma_m}$ and $\rho_{\xi}$, change with $m$.}
\label{fig:covtotal}
\end{figure}
We also look at the sum of the asymptotic posterior correlation terms involving each individual parameter estimate. For example, we define $\rho_{\mu_m}$, the asymptotic posterior correlation associated with the estimate of $\mu_m$, to be:
\begin{equation}
 \rho_{\mu_m} = |\rho_{\mu_m,\sigma_m}| + |\rho_{\mu_m,\xi}|.
\end{equation}
Figure \ref{fig:covtotal} shows how the asymptotic posterior correlation associated with each parameter varies with $m$. From Figure \ref{fig:covtotal} we see that while $\rho_{\mu_m}$ is minimised at the value of $m$ for which $\rho_{\mu_m,\sigma_m}=0$ (see Figure~\ref{fig:covmplot}), $\rho_{\sigma_m}$ and $\rho_{\xi}$ have minima at the value of $m$
for which $\rho_{\sigma_m,\xi} =0$. We denote the latter minimum by $m_1$ and the former by $m_2$. In terms of the covariance function, this can be written as:
\begin{equation}
 \text{ACov}(\sigma_{m_1}, \xi | \boldsymbol{x}) = \text{ACov}(\mu_{m_2},\sigma_{m_2} | \boldsymbol{x}) = 0,
 \label{eq:covzero}
\end{equation}
where ACov denotes the asymptotic covariance. Figure~\ref{fig:covtotal} shows that $m_2$ also minimises the total asymptotic posterior correlation in the model. \\
%Experimental evidence suggests that $m_1 < m_2$, which is consistent with the behaviour of the estimates of $m_1$ and $m_2$ with %respect to $\xi$ as shown in Figure~\ref{fig:xiplot}. \\

\begin{figure}
\centering
 \includegraphics[width=5.5cm,angle=270]{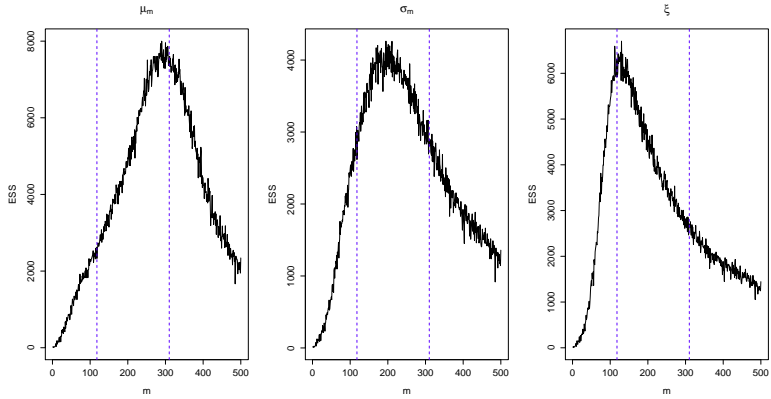}
 \caption{How ESS varies with $m$ for each parameter in ${\boldsymbol{\theta}_m}$. The blue dashed lines represent $m=m_1$ (left) and $m=m_2$ (right) in the simulated data example for 45,000 iterations of the MCMC, where $m_1$ and $m_2$ are defined by property $(\ref{eq:covzero})$. In the calculations, the sum of the autocorrelations
 were truncated when the autocorrelations in the chain drop below $0.05$.}
 \label{fig:neff}
\end{figure}

One would expect that the values of $m$ for which $\rho({\boldsymbol{\theta}_m})$ is minimised would correspond to the MCMC chain of $\boldsymbol{\theta}_m$ with good mixing properties. We examine the effective sample size (ESS) 
as a way of evaluating this objectively. ESS is a measure of the equivalent number of independent iterations that the chain represents \citep{robert2009introducing}. MCMC samples are often positively autocorrelated, and thus are less
precise in representing the posterior than if the chain was independent. The ESS of a parameter chain $\phi$ is defined as
\begin{equation}
 \text{ESS}_{\phi} = \frac{n}{1+ 2 \sum_{i=1}^{\infty} \nu_i},
\end{equation}
where $n$ is the length of the chain and $\nu_i$ denotes the autocorrelation in the sampled chain of $\phi$ at lag $i$.
In practice, the sum of the autocorrelations is truncated when $\nu_i$ drops beneath a certain level. Figure~\ref{fig:neff} shows how ESS varies with $m$ for each parameter in ${\boldsymbol{\theta}_m}$. For these data the ESS follow a pattern we found to typically occur. We see that ESS$_{\mu_m}$ is maximised at $m=m_2$ due to the near-orthogonality of $\mu_{m_2}$ with $\sigma_{m_2}$ and $\xi$. We find that ESS$_{\sigma_m}$ is maximised for $m_1 < m < m_2$, as $\sigma_{m_1}$ remains substantially positively correlated with $\mu_{m_1}$ and $\sigma_{m_2}$ is negatively correlated with $\xi$. Similarly,
ESS$_{\xi}$ is maximised at a value of $m$ close to $m_1$, but $\xi$ is negatively correlated with $\mu_{m_1}$, which explains the slight distortion. 
From these results, we postulate that a selection of $m$ in the interval $(m_1,m_2)=(118,310)$ would ensure the most rapid convergence of the MCMC chain of $\boldsymbol{\theta}_m$, thus enabling an effective sampling procedure from the joint posterior. Figure~\ref{fig:neff} shows clearly the benefits of the proposed approach. For example, ESS$_{\mu_{310}} = 7459$ and ESS$_{\mu_1} = 24$, illustrating that the former parameterisation is over $300$ times more efficient than the latter. In addition, by introducing the interval $(m_1,m_2)$, this approach gives a degree of flexibility to the choice of $m$ and giving a balance of mixing quality across the model parameters.  \\

\begin{figure}
 \centering
 \includegraphics[width=4.5cm,angle=270]{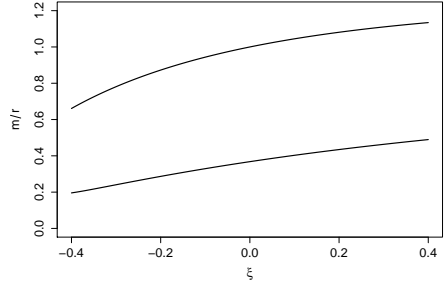}
 \caption{How $\hat{m}_1$ and $\hat{m}_2$ change as a multiple of $r$ with respect to $\hat{\xi}$: $\hat{m}_1 / r$ (bottom curve), $\hat{m}_2 / r$ (top curve).}
 \label{fig:xiplot}
\end{figure}

The quantities $m_1$ and $m_2$ can be found by numerical solution of the equations \\ $\text{ACov}(\sigma_m,\xi | \boldsymbol{x})=0$ and $\text{ACov}(\mu_m,\sigma_m | \boldsymbol{x})=0$ respectively, using the asymptotic covariance matrix of the posterior of $\boldsymbol{\theta}_m$, which is given by the inverse of the Fisher information (see Appendix~\ref{App:Stat}). Approximate analytical expressions for $m_1$ and $m_2$ can be derived using Halley's method for root-finding \citep{gander1985halley} applied to equations (\ref{eq:covzero}). This method
yields the following approximations of $m_1$ and $m_2$:
\begin{eqnarray}
  \hat{m}_1 & = & r \frac{(2\xi+1)\left(1+2\xi + (\xi+1)\log\left[\frac{2\xi+3}{2\xi+1}\right]\right)}{(2\xi+1)\left(3+2\xi - (\xi+1)\log\left[\frac{2\xi+3}{2\xi+1}\right]\right)} \label{eq:mopt1} \\
 \hat{m}_2 & = & r \frac{2\xi^2+13\xi+8}{2\xi^2+9\xi+8}.
 \label{eq:mopt2}
\end{eqnarray}
In practice, the values of $\hat{m}_1$ and $\hat{m}_2$ are estimated by using an estimate of $\xi$, such as the maximum likelihood or probability weighted moments estimates. Figure~\ref{fig:xiplot} shows how $\hat{m}_1$ and $\hat{m}_2$
change relative to $r$ for a range of $\xi$. This illustrates that for negative estimates of the shape parameter, $r$ is not a suitable candidate to be the `optimal' value of
$m$ as it is not in the range $(m_1,m_2)$. In the simulated data used in this section, although a selection of $m=r$ is reasonable, Figure~\ref{fig:neff} shows that this may not be wise if one was primarily concerned about sampling well from $\xi$, for example. In this case, $\hat{m}_2$ is relatively close to $r$, but Figure~\ref{fig:xiplot} shows that this is not the
case for models with a larger positive estimate of $\xi$. \\

A simulation study was carried out to assess the suitability of expressions $\hat{m}_1$ and $\hat{m}_2$ as approximations to $m_1$ and $m_2$ respectively. A total
of $1000$ Poisson processes were simulated with different values of $\boldsymbol{\theta}_m$. The approximations were calculated and compared with the true values
of $m_1$ and $m_2$, which were obtained exactly by numerical methods. It was found that $|\hat{m}_i - m_i| < 0.1$ for $i=1,2$ always, while $|\hat{m}_i - m_i| < 0.01$ for $78\%$ and $88.2\%$ of the time for $i=1,2$ respectively. Both quantities were compared to the performance of other approximations derived using Newton's method, which unlike Halley's method does not account for the curvature in a function. Simulations show that the root mean square errors are significantly smaller for estimates of $m_i$ using Halley's method $(0.2\%$ and $5\%$ smaller than Newton's method for $i=1,2$ respectively). A summary of the reparameterisation method is given in Algorithm~\ref{alg:repar}. \\

%\vspace{-10pt} 

\begin{algorithm}[H]
 \KwData{Threshold excesses $\boldsymbol{x}$}
 \KwResult{Samples from the posterior distribution $\pi({\boldsymbol{\theta}_k} | \boldsymbol{x})$ }
 Choose parameterisation of interest $\boldsymbol{\theta}_k$\;

 \eIf{$\boldsymbol{\theta}_k = (\mu_k, \sigma_k, \xi)$}{
  Obtain an estimate of shape parameter $\xi$ using maximum likelihood, for example\;
  Compute $\hat{m}_1$ and $\hat{m}_2$ as defined in (\ref{eq:mopt1}) and (\ref{eq:mopt2})\;
  Choose $m$ in range $(\hat{m}_1,\hat{m}_2)$\;
  }{
  Choose $m$ to be the value of $m$ that numerically solves $\rho_{\mu_m^{(0)},\sigma_m}=0$\;
  }
 Obtain MCMC samples for posterior distribution $\pi({\boldsymbol{\theta}_m} | \boldsymbol{x})$\;
 Transform to obtain samples from $\pi({\boldsymbol{\theta}_k} | \boldsymbol{x})$ using expression (\ref{eq:trans}).
 \caption{Sampling from the posterior distribution of the Poisson process model parameters $\boldsymbol{\theta}_k = (\mu_k, \sigma_k, \xi)$ or $\boldsymbol{\theta}_k = (\mu_k^{(0)}, \mu_k^{(1)}, \sigma_k, \xi)$ after reparameterising}
 \label{alg:repar}
\end{algorithm}

% \begin{table}[h!]
% \centering
% \begin{tabular}{|c|l|l|}
% \hline
%                   & \multicolumn{1}{c|}{$\hat{m}_{1}$} & \multicolumn{1}{c|}{$\hat{m}_{2}$} \\ \hline
% RMSE$_{\text{Halley}}$ &    0.0094                   &    0.0075                   \\ \cline{2-3} 
% RMSE$_{\text{Newton}}$ &    5.1610                   &     0.1472                  \\ \hline
% \end{tabular}
% \caption{Comparing the approximations for $m_1$ and $m_2$ derived from Halley's and Newton's methods in terms of Root Mean Square Error (RMSE).}
% \label{table:simstudy}
% \end{table}

\section{Choosing $m$ in the presence of non-stationarity}
\label{sec:nonstat}
In many practical applications, processes exhibit trends or seasonal effects caused by underlying mechanisms. The standard methods for modelling extremes
of non-identically distributed random variables were introduced by \citet{davison1990models} and \citet{smith1989extreme}, using a Poisson process and Generalised Pareto distribution respectively. Both approaches involve setting a constant threshold and modelling the parameters as functions of covariates. In this way,
we model the non-stationarity through the conditional distribution of the process on the covariates. We follow the Poisson process model of \citet{smith1989extreme} as the parameters are invariant to the choice of threshold if the model is appropriate. We define the covariate-dependent
parameters $\boldsymbol{\theta}_m (z) = (\mu_m (z), \sigma_m (z), \xi (z))$, for covariates $z$. Often in practice, the shape parameter $\xi$ is assumed
to be constant. A log-link is typically used to ensure positivity of $\sigma_m (z)$. \\

The process of choosing $m$ is complicated when modelling in the presence of covariates. This is partially caused by a modification of the integrated intensity measure, which becomes
\begin{equation}
 \Lambda(A) = m \int_{\mathbf{z}} {\left[1+ \xi(z)  \left(\frac{u-\mu_m (z)}{\sigma_m (z)} \right) \right]}^{-1/\xi(z)} g(z) \mathrm{d}z
 \label{eq:nsexc},
\end{equation}
where $g$ denotes the probability density function of the covariates, which is unknown and with covariate space $\mathbf{z}$.
The density term $g$ is required as the covariates associated with exceedances of the threshold $u$ are random.
In addition, the extra parameters introduced by modelling covariates increases the overall correlation in the model parameters. \\

For simplicity, we restrict our attention to the case of modelling when the location parameter is a linear function of a covariate, that is, $$\mu_m (z) = \mu_{m}^{(0)} + \mu_m^{(1)} z, \mbox{       } \sigma_m (z) = \sigma_m, \mbox{       } \xi (z) = \xi,$$
where we centre the covariate $z$, as this leads to parameters $\mu_m^{(0)}$ and $\mu_m^{(1)}$ being orthogonal. Note that the regression parameter $\mu_{m}^{(1)}$ is invariant to the choice of $m$.  A total of 233 excesses above a threshold of $u=15$ are simulated from a Poisson process model with $\mu_1^{(0)} = 75$, $\mu_1^{(1)} = 30$, $\sigma_1=15$, $\xi=-0.05$. We choose $g$ to follow an Exp$(2)$ distribution, noting that one could also choose $g$ to be the density of a covariate that is used in practice. We impose an improper Uniform prior on the regression parameter $\mu_1^{(1)}$ and set up the MCMC scheme in the same manner as in Section 3. \\

The objective remains to identify the value of $m$ that achieves near-orthogonality of the parameters of the posterior distribution. Like before, we run an MCMC sampler on $\boldsymbol{\theta}_m (z)$ and transform the samples back to the parameterisation
of interest $\boldsymbol{\theta}_k (z)$, which can be obtained as in (\ref{eq:trans}) using the relations 
\begin{eqnarray}
 \mu_{k}^{(0)} & = & \mu_{m}^{(0)} - \frac{\sigma_m}{\xi} \left(1-{\left(\frac{k}{m}\right)}^{-\xi} \right) \nonumber \\
 \mu_{k}^{(1)}& = & \mu_{m}^{(1)} \label{eq:transns}\\
 \sigma_k & = & \sigma_m {\left(\frac{k}{m}\right)}^{-\xi}. \nonumber
\end{eqnarray}

\begin{figure}[h!]
 \centering
 \includegraphics[width=8cm,angle=270]{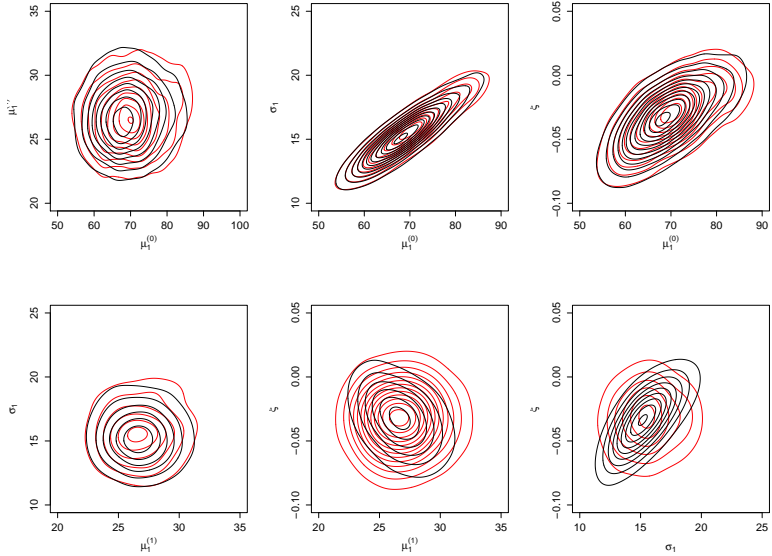}
 \caption{Contour plots of estimated posterior densities of $\boldsymbol{\theta}_1 (z)$ having sampled from the joint posterior directly (red) and having transformed using (\ref{eq:transns}) after reparameterising from $\boldsymbol{\theta}_{85} (z)$ (black). Both contours are constructed from 50,000 MCMC iterations with a burn-in of 5,000.}
 \label{fig:nsplot}
\end{figure}

The complication of the integral term in the likelihood for non-identically distributed variables means that it is no longer feasible to gain an analytical approximation for the optimal value
of $m$. A referee has suggested a possible route to obtaining such expressions for $m$ in the non-stationary case, is by building on results in \citet{attalides2015threshold} and using a non-constant threshold as in \citet{northrop2011threshold}, but as this moves away from our constant threshold case we do not pursue this. We therefore choose a value of $m$ that minimises the asymptotic posterior correlation in the model. The asymptotic posterior correlation matrix is found by inversion of the Fisher information matrix of the log-likelihood with modified integrated intensity measure~(\ref{eq:nsexc}) and normalising so that the matrix has a unit diagonal. Because of the integral term (\ref{eq:nsexc}) in the log-likelihood, the Fisher information contains various integrals that require numerical evaluation. We compute these using adaptive quadrature methods. Empirical evidence suggests that the optimal m coincides with the value of m such that $\rho_{\mu_m^{(0)}, \sigma_m} = 0$, which is similar to how $m_1$ is defined in Section 3. Using numerical methods, we identify that this corresponds to a value of $m=85$ for the simulated data example. Figure \ref{fig:nsplot} shows contour plots of estimated posterior densities of $\boldsymbol{\theta}_1 (z)$, comparing the sampling from directly estimating the posterior $\boldsymbol{\theta}_1 (z)$
with that from transforming the samples from the estimated posterior of $\boldsymbol{\theta}_{m} (z)$ to give a sample from the posterior of $\boldsymbol{\theta}_1 (z)$.
From this figure, we see that the reparameterisation improves the sampling from the posterior $\boldsymbol{\theta}_1 (z)$. \\

% \begin{figure}
%  \centering
%  \includegraphics[width=12cm]{nsindcor.eps}
%  \caption{Total correlations associated with each parameter estimated using simulated data: $\mu_0$ (black), $\mu_1$ (red), $\sigma$ (green), $\xi$ (blue)}
%  \label{fig:indnscor}
% \end{figure}
\begin{figure}[h!]
 \centering
 \includegraphics[width=8cm,angle=270]{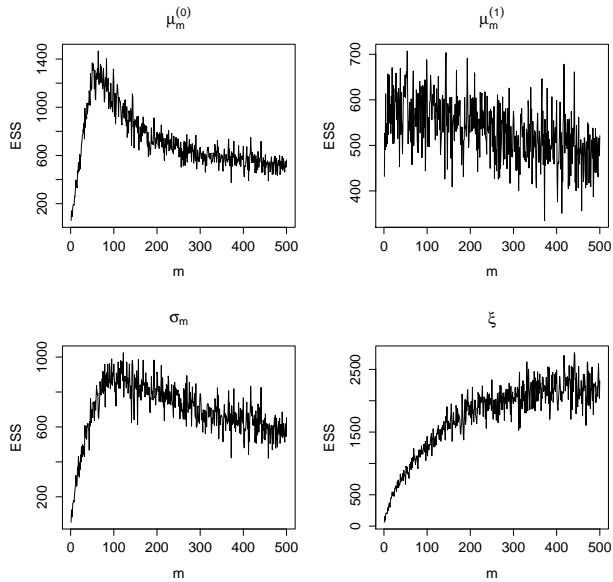}
 \caption{Effective sample size of each parameter chain of the MCMC procedure.}
 \label{fig:essns}
\end{figure}

We again inspect the effective sample size for each parameter as a way of comparing the efficiency of the MCMC under different parameterisations.
Figure~\ref{fig:essns} shows how the effective sample size varies with $m$ for each parameter. This figure shows how the quality of mixing is approximately maximised in $\mu_m^{(0)}$ for the value of $m$ that minimises the asymptotic posterior correlation. Mixing for $\mu_m^{(1)}$ is consistent
across all values of $m$. Interestingly, mixing in $\xi$ increases as the value of $m$ increases. Without a formal measure for the quality of mixing across the parameters,
it is found that, when averaging the effective sample size over the number of parameters, the ESS is stable with respect to $m$ in the interval spanning from the value of $m$ such that $\rho_{\mu_m^{(0)},\sigma_m}=0$ and the value of $m$ such that $\rho_{\sigma_m,\xi}=0$, like in
Section~\ref{sec:choose}. For a summary of how the reparameterisation method can be used in the presence of non-stationarity, see Algorithm~\ref{alg:repar}. \\

\section{Case study: Cumbria rainfall}
\label{sec:cumbria}
In this section, we present a study as an example of how this reparameterisation method can be used in practice. In particular, we analyse data taken from the Met Office
UKCP09 project, which contains daily baseline averages of surface rainfall observations, measured in millimetres, in 25km $\times$ 25km grid cells across the United Kingdom in the period 1958-2012. In this analysis, we focus on a grid cell
in Cumbria, which has been affected by numerous flood events in recent years, most notably in 2007, 2009 and 2015. In particular, the December 2015 event resulted in an estimated $\pounds 5$ billion worth of damage, with rain gauges reaching unprecedented levels.
Many explanations have been postulated for the seemingly increased rate of flooding in the North West of England, including climate change, natural climate variability or a combination of both. The baseline average data for the flood events in December 2015 are not yet available, but this event is widely regarded as being more extreme
than the event in November 2009, the levels of which were reported at the time to correspond to return periods of greater than $100$ years. We focus our analysis on the 2009 event, looking in particular at how a phase of climate variability, in the form of the North Atlantic Oscillation (NAO) index, can have a significant
impact on the probability of an extreme event occurring in any given year.\\

\begin{figure}[h!]
 \centering
 \includegraphics[width=8cm,angle=270]{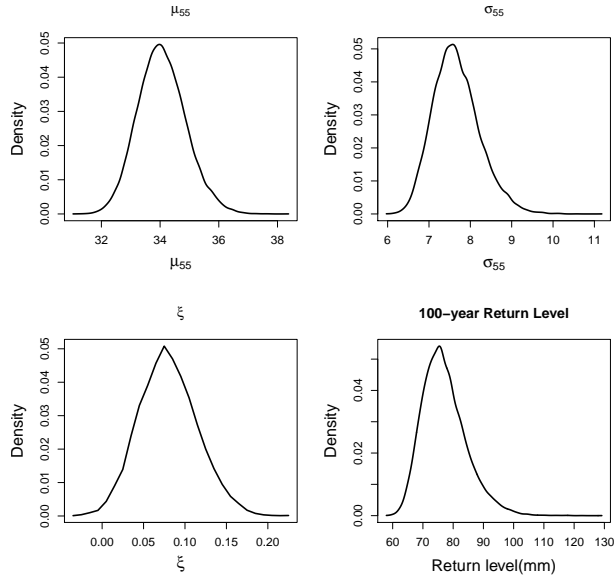}
 \caption{Estimated posterior densities of $\mu_{55}$, $\sigma_{55}$, $\xi$ and the 100-year return level.}
 \label{fig:post_dens}
\end{figure}

Rainfall datasets on a daily scale are commonly known to exhibit a degree of serial correlation. Analysis of autocorrelation and partial autocorrelation plots indicates that rainfall on a day is dependent on the rainfall of the previous five days. In addition, the data may exhibit seasonal effects. However, while serial dependence affects the effective sample size of a dataset, it does not affect correlations between parameters, and is thus unlikely to influence the choice of $m$. For the purposes of illustrating our method, we initially make the assumption that the rainfall observations are iid and proceed with the method outlined in Section~\ref{sec:choose}. We wish to obtain information
about the parameters corresponding to the distribution of annual maxima, i.e. $\boldsymbol{\theta}_{55}$.
\begin{figure}[h!]
 \centering
 \includegraphics[width=6cm,angle=270]{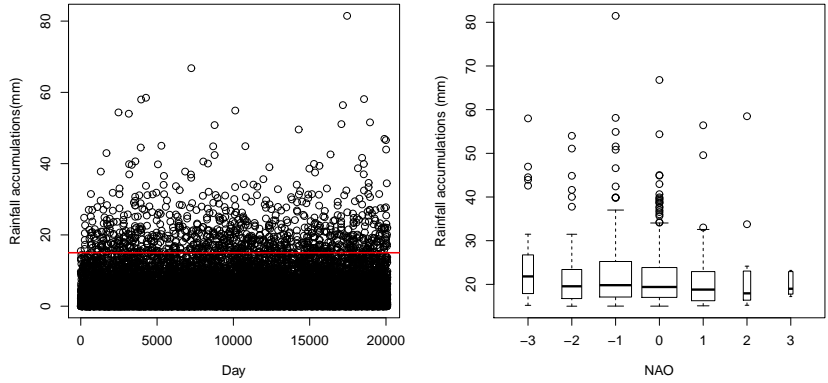}
 \caption{(Left) Daily rainfall observations in the Cumbria grid cell in the period 1958-2012. The red line represents the extreme value threshold of $u=15$. (Right) Boxplots of rainfall above $u$ against the corresponding monthly NAO index.}
 \label{fig:cumbria}
\end{figure}
Standard threshold diagnostics \citep{coles2001introduction} indicate a threshold of $u=15$ is appropriate, which corresponds to the $95.6\%$ quantile of the data. There are $r=880$ excesses above $u$ (see Figure~\ref{fig:cumbria}).
We obtain bounds $m_1$ and $m_2$, then choose a value of $m$, with $m_1 < m < m_2$, that will achieve near-orthogonality of the Poisson process model parameters to improve MCMC sampling from the joint posterior distribution.
We obtain $\hat{\xi}=0.087$ using maximum likelihood when $m=r$, which we use to obtain approximations for $m_1$ and $m_2$ as in (\ref{eq:mopt1}) and (\ref{eq:mopt2}). From this, we obtain $\hat{m}_1 \approx 351$ and $\hat{m}_2\approx915$.
We checked that $\hat{m}_1$ and $\hat{m}_2$ represent good approximations by solving equations (\ref{eq:covzero}) to obtain $m_1 = 350.82$ and $m_2 = 914.96$.
Since $r=880$ is contained in the interval $(m_1,m_2)$, we choose $m=r$.
We run an MCMC chain for $\boldsymbol{\theta}_{880}$ for 50,000 iterations, discarding the first 1,000 samples as burn-in.
We transform the remaining samples using the mapping in (\ref{eq:trans}), where $k=55$, to obtain samples from the joint posterior of $\boldsymbol{\theta}_{55}$. The estimated posterior density for each parameter is shown in
Figure~\ref{fig:post_dens}. \\

To estimate probabilities of events beyond the range of the data, we can use the estimated parameters to estimate extreme quantiles of the annual maximum distribution.
The quantity $y_N$, satisfying:
\begin{equation}
 1/N = 1-G(y_N),
 \label{eq:retlev}
\end{equation}
is termed the $N$-year return level, where $G$ is defined as in expression (\ref{eq:gev}). The level $y_N$ is expected to be exceeded on average once every $N$ years. By inverting (\ref{eq:retlev}) we get:
\begin{equation}
  y_{N} =
 \left \{ \begin{array}{lr}
\mu_{55} - \frac{\sigma_{55}}{\xi}[1-{\{-\log(1-1/N)\}}^{-\xi}] & \mbox{for   } \xi \neq 0 \\
\mu_{55} - \sigma_{55}\log\{-\log(1-1/N)\} & \mbox{for   } \xi = 0.  \\ 
\end{array} \right. 
\label{eq:retformula}
\end{equation}

The posterior density of the 100-year return level in Figure~\ref{fig:post_dens} is estimated by inputting the MCMC samples of the model parameters into expression (\ref{eq:retformula}). \\

% \begin{figure}[h!]
%  \centering
%  \includegraphics[width=10cm]{ret_levels_iid.eps}
%  \caption{Estimated return levels for rainfall data.}
%  \label{fig:ret_lev}
% \end{figure}

We use the same methodology to explore the effect of the monthly NAO index on the probability of extreme rainfall levels in Cumbria.
The NAO index describes the surface sea-level pressure difference between the Azores High and the Icelandic Low. The low frequency variability of the monthly scale is chosen
to represent the large scale atmospheric processes affecting the distribution of wind and rain. In the UK, a positive NAO index is
associated with cool summers and wet winters, while a negative NAO index typically corresponds to cold winters, pushing the North Atlantic storm track
further south to the Mediterranean region \citep{hurrell2003overview}. In this analysis,
we incorporated the effect of NAO by introducing it as a covariate in the location parameter. The threshold of $u=15$ was retained for this analysis. \\

\begin{figure}[h!]
 \centering
 \includegraphics[width=8cm,angle=270]{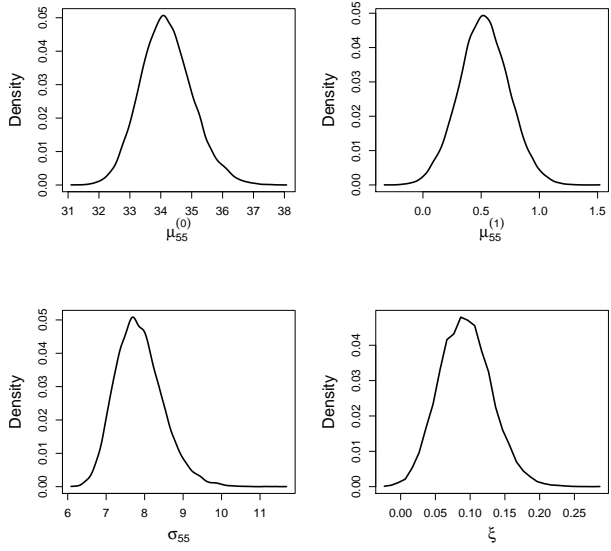}
 \caption{Estimated posterior densities of $\mu^{(0)}_{55}$, $\mu^{(1)}_{55}$, $\sigma_{55}$ and $\xi$.}
 \label{fig:ns_dens}
\end{figure}

To obtain the value of $m$ that minimises the overall correlation in the model, we solve numerically the equation
$\rho_{\mu_m^{(0)},\sigma_m}=0$, following the reasoning in Section~\ref{sec:nonstat}. We obtain a kernel density
estimate of the NAO covariate, which represents $g$ as defined in expression (\ref{eq:nsexc}). We use this to obtain maximum posterior mode estimates $\hat{\boldsymbol{\theta}}_r $. These quantities are substituted into the Fisher information matrix. The matrix is then inverted numerically to estimate $m = 920$. This represents a slight deviation
from $\hat{m}_2$ estimated during the iid analysis. We would expect this as the covariate effect is small, as shown in Figure~\ref{fig:ns_dens}. This example illustrates the benefit of numerically solving for $m$ when modelling non-stationarity, as the range $(m_1,m_2)$ estimated analytically during the iid analysis no longer contain the optimal value of $m$. \\

We run an MCMC chain for $\boldsymbol{\theta}_{920}$ for 50,000 iterations before discarding the first 5,000 samples as burn-in.
We transform the remaining MCMC samples to the annual maximum scale using the mapping in (\ref{eq:transns}) where $k=55$. Figure~\ref{fig:ns_dens} indicates that NAO has a significantly positive effect on the location parameter, as almost all posterior mass is distributed with $\mu_{55}^{(1)} > 0$.  \\

% \begin{figure}[h!]
%  \centering
%  \includegraphics[width=10cm]{ret_lev_NAO.eps}
%  \caption{The 100-year return level changing with NAO, with the full line representing the posterior and the two dashed lines representing $95\%$ credible intervals.}
%  \label{fig:retlev}
% \end{figure}

\begin{figure}[h!]
 \centering
 \includegraphics[width=8cm]{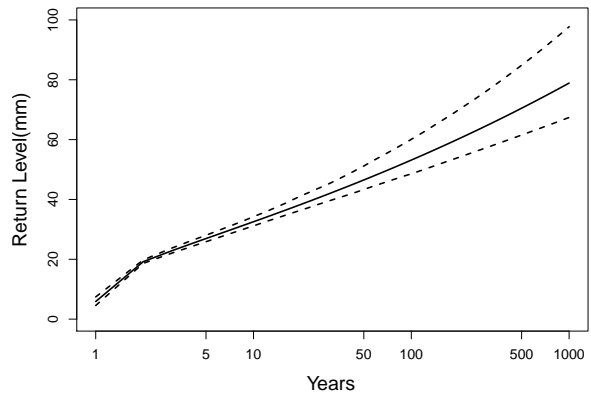}
 \caption{Return levels corresponding to November maxima. The full line represents the posterior mean and the two dashed lines representing $95\%$ credible intervals.}
 \label{fig:condretlev}
\end{figure}

% In terms of return levels, Figure~\ref{fig:retlev} shows
% that positive NAO indices give larger return level estimates than negative NAO indices, which is what we would expect in the United Kingdom. 
We wish to estimate return levels relating to the November 2009 flood event, which is represented by a value of $51.6$mm in the dataset. Return levels corresponding to the distribution of November maxima are shown in Figure~\ref{fig:condretlev}. We can also use the predictive distribution in order to account for both parameter
uncertainty and randomness in future observations \citep{coles1996bayesian}. On the basis of threshold excesses $\mathbf{x}=(x_1, \hdots, x_n)$, the predictive distribution of a future November maximum $M$ is:

\begin{equation}
\Pr\{M\leq y|\mathbf{x}\} = \int_{\boldsymbol{\theta}_{55}} \Pr\{M\leq y|\boldsymbol{\theta}_{55}\}\pi(\boldsymbol{\theta}_{55}|\mathbf{x}) \mathrm{d} \boldsymbol{\theta}_{55},
\label{eq:predict}
\end{equation}
 where $\Pr\{M\leq y|\boldsymbol{\theta}_{55}\} =$
\begin{align*}
  \begin{cases} \exp \left\{- \frac{1}{12} { \left[1+\xi \left( \frac{y-(\mu^{(0)}_{55} + \mu^{(1)}_{55} z)}{\sigma_{55}} \right) \right] }_{+}^{-1/\xi} \right\}  & \mbox{where } z\mbox{ is known} \vspace{10pt} \\ 
 \exp \left\{- \frac{1}{12} \displaystyle\int_{z} { \left[1+\xi \left( \frac{y-(\mu^{(0)}_{55} + \mu^{(1)}_{55} z)}{\sigma_{55}} \right) \right] }_{+}^{-1/\xi} g_{N} (z) \mathrm{d}z \right\}  & \mbox{where } z\mbox{ is unknown}, \end{cases}
\end{align*}
where $g_N$ is the density of NAO in November and the integral is evaluated numerically using adaptive quadrature methods. The integral in (\ref{eq:predict}) can be approximated using a Monte Carlo summation over the samples from the joint posterior of $\boldsymbol{\theta}_{55}$. From this, we estimate the predictive probability of an event exceeding 51.6 in a typical November is $0.0112$, with a $95\%$ credible interval of $(0.0063,0.0185)$, which corresponds to an $89$-year event, $(54,158)$. For November 2009, when
an NAO index of $-0.02$ was measured, the probability of such an event was $0.0111$, $(0.0062,0.0184)$, corresponding to a 90-year event, $(54,161)$. For the maximum observed value of NAO in November, with $\text{NAO}=3.04$, the predictive probability of such an event
is $0.0132$, $(0.0073,0.0214)$, which corresponds to a 75-year flood event, $(47,136)$. This illustrates that the impact that different phases of climate variability can have on the probabilities of extreme events is slight but potentially important.

\newpage
\section*{Appendix}
\appendix
%\section{Probability density of a non-homogenous Poisson process model for extremes}
%\label{App:Dens}
%Consider a two-dimensional non-homogeneous Poisson process with intensity $\lambda(t,x)$ on the set $A=[0,1] \times (u,\infty)$, for some finite $u$. 
%For a non-stationary limiting point process, the intensity function $\lambda$ is of the form
%\begin{equation*}
% \lambda(t,x) = \frac{1}{\sigma} {\left\{1+ \xi \left(\frac{x-\mu}{\sigma}\right)\right\}}^{-1/{\xi} - 1}, 
% \end{equation*}
%for model parameters $\theta = (\mu, \sigma, \xi)$. \\
%Let $N(A)$ be the number of points of the Poisson process in the set
%$A$. A key property of a Poisson process is that
%\[
%N(A) \sim \text{Poisson}(\Lambda(A)), 
%\]
%where $\Lambda(A)$ is the integrated intensity function
%\begin{equation*}
%\Lambda(A) = \int_{u}^{\infty} \lambda(t,x) dx. 
%\end{equation*}
%The density of points in the set $A$ at the point $(t,x)$ is defined as
%\begin{equation*}
% f(t,x) = \frac{\lambda(t,x)}{\Lambda(A)}, \mbox{                 for             } t \in [0,1], x \in [u,\infty) 
%\end{equation*}
%For iid random variables,
%\[  f(x) = \frac{{\left[1+ \xi \left(\frac{x-\mu}{\sigma}\right)\right]}^{-1/{\xi} -1}}{  {\left[1+ \xi \left(\frac{u-\mu}{\sigma}\right) \right]}^{-1/{\xi}}} \]
\section{Proof: $\hat{\mu}_r = u$ when $m=r$}
\label{App:gp}
We can write the full likelihood for parameters $\boldsymbol{\theta}_r$ given a series of excesses $\{x_i\}$ above a threshold $u$ as:
$$ L(\boldsymbol{\theta}_r) = L_1 \times L_2,$$
where $L_1$ is the Poisson probability of $r$ exceedances of $u$ and $L_2$ is the joint density of these $r$ exceedances, so that:
\begin{eqnarray*}
 L_1 &=& \frac{1}{r!} {\left\{r {\left[1+ \xi \left(\frac{u-\mu_r}{\sigma_r}\right) \right]}_{+}^{-1/{\xi}} \right\}}^r \exp\left\{-r{\left[1+ \xi \left(\frac{u-\mu_r}{\sigma_r}\right) \right]}_{+}^{-1/{\xi}}\right\}, \\
 L_2 &=& \prod_{i-1}^{r} \frac{1}{\sigma_r} {\left[1+ \xi \left(\frac{x_i-\mu_r}{\sigma_r}\right) \right]}_{+}^{-1/{\xi} - 1} {\left[1+ \xi \left(\frac{u-\mu_r}{\sigma_r}\right) \right]}_{+}^{1/{\xi}}.
\end{eqnarray*}
By defining $\Lambda =  {\left[1+ \xi \left(\frac{u-\mu_r}{\sigma_r}\right) \right]}_{+}^{-1/{\xi}} $ and $\psi_u = \sigma_r + \xi (u-\mu_r)$ we can reparameterise
the likelihood in terms of $\boldsymbol{\theta}^{*} = (\Lambda,\psi_u, \xi)$ to give:
\begin{eqnarray*}
L(\boldsymbol{\theta}^{*}) &\propto& \Lambda^r \exp\left\{-r \Lambda\right\}\prod_{i=1}^r \frac{1}{\psi_u - \xi(u-\mu_r)} {\left[ \frac{\psi_u + \xi(x_i -u)}{\psi_u - \xi(u-\mu_r)}\right]}_{+}^{-1/\xi - 1} {\left[ \frac{\psi_u}{\psi_u - \xi(u-\mu_r)} \right]}_{+}^{1/\xi}  \\
  &=& \Lambda^r \exp\left\{-r \Lambda\right\}\prod_{i=1}^r \frac{1}{\psi_u} {\left[1+ \xi \left(\frac{x_i-u}{\psi_u}\right) \right]}_{+}^{-1/{\xi} -1 }.
\end{eqnarray*}
Taking the log-likelihood and maximising with respect to $\Lambda$, we get:
\begin{eqnarray*}
l(\boldsymbol{\theta}^{*}) &:=& \log L(\boldsymbol{\theta}^{*}) = r \log \hat{\Lambda} - r \hat{\Lambda} -r \log \psi_u - \left(\frac{1}{\xi} + 1\right) \sum_{i=1}^r \log{\left[1+ \xi \left(\frac{x_i-u}{\psi_u}\right) \right]}_{+}  \\
\frac{\partial l}{\partial \Lambda} &=&   \frac{r}{\hat{\Lambda}} - r  = 0, 
\end{eqnarray*}
which gives $\hat{\Lambda} = 1$. Then, by the invariance property of maximum likelihood estimators, $\hat{\mu}_r = u$, and using the identity for $\psi_u$, we get $\hat{\sigma}_r= \hat{\psi}_u$.
%\begin{align*}
% \hat{\Lambda} = {\left[1+ \hat{\xi} \left(\frac{u-\hat{\mu}_r}{\hat{\sigma}_r}\right) \right]}_{+}^{-1/{\hat{\xi}}} & =  1 \\
%\frac{u-\hat{\mu}_r}{\hat{\sigma}_r} & =  0. \\
% \hat{\mu}_r & = u.
%\end{align*}
Because the $\xi$-dependent term in the log-likelihood is identical to that in a GP log-likelihood, the maximum likelihood estimators of the two models coincide.

%$\hat{\Lambda}=1$, which in turn means that $\hat{\mu}_r = u$.

\newpage
\section{Derivation of prior for inference on $\boldsymbol{\theta}_m$}
\label{App:prior}
We define a joint prior on the parameterisation of interest $\boldsymbol{\theta}_k$. However, as we are making inference for the `optimal' parameterisation $\boldsymbol{\theta}_m$, we must derive the prior for $\boldsymbol{\theta}_m$. We can calculate the prior density of $\boldsymbol{\theta}_m$ by using the density method for one-to-one bivariate transformations. Inverting (\ref{eq:trans}) to get expressions for $\mu_m$ and $\sigma_m$, i.e. 
\begin{eqnarray}
 \mu_m & = & \mu_k - \frac{\sigma_k}{\xi} \left(1-{\left(\frac{m}{k}\right)}^{-\xi} \right) = g_1 (\mu_k, \sigma_k) \nonumber \\
 \sigma_m & = & \sigma_k {\left(\frac{m}{k}\right)}^{-\xi} = g_2 (\mu_k, \sigma_k)  \nonumber \label{eq:transapp},
\end{eqnarray}
we can use this transformation to calculate the prior for $\boldsymbol{\theta}_m$.
\begin{align*}
\pi(\boldsymbol{\theta}_m) & = \pi(\mu_m,\sigma_m,\xi) \\
						   & = \pi(\mu_k,\sigma_k,\xi) {| \det {J} |}_{\mu_k = g_{1}^{-1} (\mu_m, \sigma_m) , \sigma_k = g_{2}^{-1} (\mu_m, \sigma_m), \xi = \xi}, \\
\intertext{where} 
 \det J &= \left| \begin{array}{ccc}
\frac{\partial \mu_m}{\partial \mu_k}& \frac{\partial \mu_m}{\partial \sigma_k} & \frac{\partial \mu_m}{\partial \xi} \\
\frac{\partial \sigma_m}{\partial \mu_k}& \frac{\partial \sigma_m}{\partial \sigma_k} & \frac{\partial \sigma_m}{\partial \xi} \\
\frac{\partial \xi}{\partial \mu_k}& \frac{\partial \xi}{\partial \sigma_k} & \frac{\partial \xi}{\partial \xi} \end{array} \right| \\
& = \left| \begin{array}{ccc}
\frac{\partial \mu_m}{\partial \mu_k}& \frac{\partial \mu_m}{\partial \sigma_k} & \frac{\partial \mu_m}{\partial \xi} \\
0& \frac{\partial \sigma_m}{\partial \sigma_k} & \frac{\partial \sigma_m}{\partial \xi} \\
0& 0 & \frac{\partial \xi}{\partial \xi} \end{array} \right| \\
& = \frac{\partial \sigma_m}{\partial \sigma_k} \frac{\partial \xi}{\partial \xi} \\
& = {\left(\frac{m}{k}\right)}^{-\xi}.
\end{align*}
Therefore,  $\pi(\boldsymbol{\theta}_m) = {\left(\frac{m}{k}\right)}^{-\xi} \pi(\boldsymbol{\theta}_k).$
%Since, in our case, the prior on $\boldsymbol{\theta}_k$ is Uniform, \newline $\pi(\boldsymbol{\theta}_m) = {\left(\frac{m}{k}%\right)}^{-\xi}$.

\section{Fisher information matrix calculations for iid random variables}
\label{App:Stat}
The log-likelihood of the Poisson process model with parameterisation $\boldsymbol{\theta}_m= (\mu_m,\sigma_m, \xi)$ can be expressed as
\[
l(\boldsymbol{\theta}_m) = -m {\left[1+ \xi \left(\frac{u-\mu_m}{\sigma_m}\right) \right]}_{+}^{-1/{\xi}} - r \log \sigma_m - \left(\frac{1}{\xi} +1\right)\sum_{j=1}^{r} \log{\left[1+ \xi \left(\frac{x_j - \mu_m}{\sigma_m}\right)\right]}_{+},
\]
where $r$ is the number of exceedances of $X$ above the threshold $u$. For simplicity, we drop the ${[\cdot]}_{+}$ subscript in subsequent calculations. In order to produce analytic expressions for the asymptotic covariance matrix, we must evaluate the observed information matrix $\hat{I}(\boldsymbol{\theta}_m)$. For simplicity, we define
$v_m = \frac{u-\mu_m}{\sigma_m}$ and $z_{j,m} = \frac{x_j - \mu_m}{\sigma_m}$.

\begin{eqnarray*}
\frac{\partial^2 l}{\partial \mu_m^2} &=& -\frac{m(\xi+1)}{\sigma_m^2} \rate^{-1/\xi - 2} + \frac{\xi(\xi+1)}{\sigma_m^2} \sumr \xrate^{-2}, \\
\frac{\partial^2 l}{\partial \sigma_m^2} &=& \frac{2m}{\sigma_m^2} \rate^{-1/\xi -1} \umu - \frac{m(\xi+1)}{\sigma_m^2} \rate^{-1/\xi -2} \umu^2 + \frac{r}{\sigma_m^2} - \frac{2(\xi+1)}{\sigma_m^2} \sumr \xrate^{-1} \xmu \\ & & + \frac{\xi(\xi+1)}{\sigma_m^2} \sumr \xmu^2 \xrate^{-2}, \\
\frac{\partial^2 l}{\partial \xi^2} &=& -m \rate^{-1/\xi} \left[ \frac{1}{\xi}\umu^2 \rate^{-2} - \frac{2}{\xi^3}\log\rate \right.\\ & & +\left.\frac{2}{\xi^2} \rate^{-1} \umu + {\left(\frac{1}{\xi^2}\log \rate - \frac{1}{\xi} \rate^{-1} \umu \right)}^2 \right]  \\ & &- \frac{2}{\xi^3} \sumr \log\xrate + \frac{2}{\xi^2} \sumr \xrate^{-1} \xmu + \frac{\xi+1}{\xi} \sumr \xrate^{-2} \xmu^2, \\
\frac{\partial^2 l}{\partial \mu_m \partial \sigma_m} & = & \frac{m}{\sigma_m^2} \rate^{-1/\xi - 1}  - \frac{m(\xi+1)}{\sigma_m^2} \rate^{-1/\xi -2} \umu \\ & &- \frac{\xi+1}{\sigma_m^2} \sumr \xrate^{-1} + \frac{\xi(\xi+1)}{\sigma_m^2} \sumr \xrate^{-2} \xmu , \\
\frac{\partial^2 l}{\partial \mu_m \partial \xi} & = & -\frac{m}{\sigma_m} \left[ \frac{1}{\xi^2} \rate^{-1/\xi - 1} \log \rate - \frac{\xi+1}{\xi} \rate^{-1/\xi - 2} \umu \right] + \frac{1}{\sigma_m} \sumr \xrate^{-1} \\ & & - \frac{\xi+1}{\sigma_m} \sumr \xrate^{-2} \xmu, \\
\frac{\partial^2 l}{\partial \sigma_m \partial \xi} & = & -\frac{m}{\sigma_m} \umu \left[ \frac{1}{\xi^2} \rate^{-1/\xi - 1} \log \rate - \frac{\xi+1}{\xi} \rate^{-1/\xi - 2} \umu \right] + \frac{1}{\sigma_m} \sumr \xrate^{-1} \xmu \\ & & - \frac{\xi+1}{\sigma_m} \sumr \xrate^{-2} \xmu^2
\end{eqnarray*}
To obtain the Fisher information matrix, we take the expected value of each term in the observed information with respect to the probability density of points of a Poisson process. Let $Z= \frac{X-\mu_m}{\sigma_m}$, and $R$ be a random variable denoting the number of excesses of $X$ above $u$. The density of points in the set $A_u$ can de defined by
\[ f(x) = \frac{\lambda(x)}{\Lambda(A_u)} = \frac{{\left[1+ \xi z \right]}^{-1/\xi - 1}}{\rate^{-1/\xi}}, \]
where $\lambda$ is a function denoting the rate of exceedance.
 Then, for example,
\begin{eqnarray*}
 \mathbb{E}_{Z,R}\left\{ \sumR \xrate^{-2} \right\} & = & \mathbb{E}_R \mathbb{E}_{Z|R} \left\{ \sumR \xrate^{-2} \right\} \\
 & = & \mathbb{E}_R \left\{ R  \mathbb{E}_Z \left\{ {\left[1+ \xi Z \right]}^{-2} \right\} \right\} \\
 & = & \mathbb{E}_R \left\{ R \rate^{1/\xi} \int_{v_m}^{\infty} \zrate^{-1/\xi - 3} \mathrm{d}z \right\}\\
 & = & \frac{m}{2\xi+1} \rate^{-1/\xi - 2}
\end{eqnarray*}
Following this process, we can write the Fisher information matrix $I(\boldsymbol{\theta}_m)$ as:
\begin{eqnarray*}
 \mathbb{E}\left\{ -\frac{\partial^2 l}{\partial \mu_m^2} \right\} &=& \frac{m(\xi+1)}{\sigma_m^2} \rate^{-1/\xi - 2} - \frac{m\xi(\xi+1)}{(2\xi+1)\sigma_m^2} \rate^{-1/\xi-2}, \\
\mathbb{E}\left\{-\frac{\partial^2 l}{\partial \sigma_m^2} \right\}&=& -\frac{2m}{\sigma_m^2} \rate^{-1/\xi -1} \umu + \frac{m(\xi+1)}{\sigma_m^2} \rate^{-1/\xi -2} \umu^2 - \frac{r}{\sigma_m^2} + \\
& & \frac{2m}{\sigma_m^2} \rate^{-1/\xi -1} \left[1+(\xi+1)\umu\right] - \\
 & & \frac{m\xi}{(2\xi+1)\sigma_m^2} \rate^{-1/\xi - 2} \left[(2\xi^2+3\xi+1)\umu^2 + (4\xi+2)\umu + 2\right], \\
 \mathbb{E}\left\{-\frac{\partial^2 l}{\partial \xi^2} \right\} &=& m \rate^{-1/\xi} \left[ \frac{1}{\xi}\umu^2 \rate^{-2} - \frac{2}{\xi^3}\log\rate + \right.\\ 
 & & \left.\frac{2}{\xi^2} \rate^{-1} \umu + {\left(\frac{1}{\xi^2}\log \rate - \frac{1}{\xi} \rate^{-1} \right)}^2 \right] +  \\ 
 & & \frac{2}{\xi^3} \rate^{-1/\xi}\left[\xi + \log\rate\right] -\frac{2m}{(\xi+1)\xi^2} \rate^{-1/\xi-1} \left[1+(\xi+1)\umu\right] -\\ 
& &  \frac{m}{\xi(2\xi+1)} \rate^{-1/\xi - 2} \left[(2\xi^2+3\xi+1)\umu^2 + (4\xi+2)\umu + 2\right], \\
\mathbb{E}\left\{-\frac{\partial^2 l}{\partial \mu_m \partial \sigma_m} \right\} & = & \frac{m(\xi+1)}{\sigma_m^2} \rate^{-1/\xi - 2} \umu - \frac{m\xi}{(2\xi+1)\sigma_m^2} \rate^{-1/\xi - 2} \left[1+(2\xi+1)\umu\right]  , \\
\mathbb{E}\left\{-\frac{\partial^2 l}{\partial \mu_m \partial \xi} \right\} & = & \frac{m}{\sigma_m} \left[ \frac{1}{\xi^2} \rate^{-1/\xi - 1} \log \rate -  \frac{\xi+1}{\xi} \rate^{-1/\xi - 2} \umu \right] - \\
& & \frac{m}{\sigma_m(\xi+1)} \rate^{-1/\xi -1} + \frac{m}{\sigma_m(2\xi+1)} \rate^{-1/\xi - 2} \left[1+(2\xi+1) \umu \right], \\
\mathbb{E}\left\{-\frac{\partial^2 l}{\partial \sigma_m \partial \xi} \right\} & = & \frac{m}{\sigma_m} \umu \left[ \frac{1}{\xi^2} \rate^{-1/\xi - 1} \log \rate + \frac{\xi+1}{\xi} \rate^{-1/\xi - 2} \umu \right] - \\ & & \frac{m}{\sigma_m(\xi+1)} \rate^{-1/\xi - 1} \left[1+(\xi+1)\umu\right] +\\ & &  \frac{m}{\sigma_m(2\xi+1)} \rate^{-1/\xi - 2} \left[(2\xi^2+3\xi+1)\umu^2 + (4\xi+2)\umu + 2\right].  
\end{eqnarray*}
By inverting the Fisher information matrix using a technical computing tool like Wolfram Mathematica, making the substitution $r = m \rate^{-1/\xi}$, the expected number of exceedances, and using the mapping in (\ref{eq:trans}), we can get expressions for asymptotic posterior covariances.
\begin{eqnarray*}
 \text{ACov}(\mu_m,\xi) & =& \frac{1}{\xi ^2 r}(\xi +1) \sigma_m   \left(\frac{r}{m}\right)^{-\xi } \left(\xi  (\xi +1) \left(\frac{r}{m}\right)^{\xi } \log \left(\frac{r}{m}\right)-(2 \xi +1) \left(\left(\frac{r}{m}\right)^{\xi }-1\right)\right) \\
 \text{ACov}(\mu_m,\sigma_m) & = & \frac{1}{ \xi ^2 r}\sigma_{m}^{2}  \left(\frac{r}{m}\right)^{-\xi } \left(\left(\frac{r}{m}\right)^{\xi } \left((\xi +1) \log \left(\frac{r}{m}\right) \left((\xi +1) \xi  \log \left(\frac{r}{m}\right)-3 \xi -1\right)+ \right. \right. \\ & &  \xi  (\xi  (\xi +2)+3)+1 \Big)+(\xi +1) (2 \xi +1) \left(\log \left(\frac{r}{m}\right)-1\right)\bigg) \\
 \text{ACov}(\sigma_m,\xi) & = & \frac{1}{r} (\xi +1) \sigma_m   \left((\xi +1) \log \left(\frac{r}{m}\right)-1\right)
\end{eqnarray*}
When $m=r$, $\text{ACov}(\mu_m,\xi)= 0$. In addition, the $m$ for which $\text{ACov}(\mu_m,\sigma_m) = 0$ coincides with the value of $m$ that minimises $\rho_{\boldsymbol{\theta}_m}$ as defined in (\ref{eq:totcor}). This root can easily be found numerically, but an analytical approximation can be calculated using a
one-step Halley's method. By using $m=r$ as the initial seed, and using the formula:
\begin{equation*}
 x_{n+1} = x_n - \frac{f(x_n)}{f'(x_n) - \frac{f(x_n) f''(x_n)} {2 f'(x_n)}}
\end{equation*}
we get the expression (\ref{eq:mopt2}) for $\hat{m}_2$ after one step. The quantity for $\hat{m}_1$, given by expression (\ref{eq:mopt1}) requires two iterations of this method.

\section*{Acknowledgements}
We gratefully acknowledge the support of the EPSRC funded EP/H023151/1 STOR-i Centre for Doctoral Training, the Met Office and EDF Energy. We extend our thanks to Jenny Wadsworth of Lancaster University, Simon Brown of the Met Office and two referees 
for very helpful comments. We also thank the Met Office for the rainfall data.
\bibliographystyle{spbasic}      % basic style, author-year citations
\bibliography{ppbiblio}   % name your BibTeX data base

\begin{thebibliography}{22}
\providecommand{\natexlab}[1]{#1}
\providecommand{\url}[1]{{#1}}
\providecommand{\urlprefix}{URL }
\expandafter\ifx\csname urlstyle\endcsname\relax
  \providecommand{\doi}[1]{DOI~\discretionary{}{}{}#1}\else
  \providecommand{\doi}{DOI~\discretionary{}{}{}\begingroup
  \urlstyle{rm}\Url}\fi
\providecommand{\eprint}[2][]{\url{#2}}

\bibitem[{Attalides(2015)}]{attalides2015threshold}
Attalides N (2015) Threshold-based extreme value modelling. PhD thesis, UCL
  (University College London)

\bibitem[{Chavez-Demoulin and Davison(2005)}]{chavez2005generalized}
Chavez-Demoulin V, Davison AC (2005) Generalized additive modelling of sample
  extremes. Journal of the Royal Statistical Society: Series C (Applied
  Statistics) 54(1):207--222

\bibitem[{Coles(2001)}]{coles2001introduction}
Coles SG (2001) An Introduction to Statistical Modeling of Extreme Values.
  Springer

\bibitem[{Coles and Tawn(1996)}]{coles1996bayesian}
Coles SG, Tawn JA (1996) A {B}ayesian analysis of extreme rainfall data.
  Applied Statistics 45:463--478

\bibitem[{Cox and Reid(1987)}]{cox1987parameter}
Cox DR, Reid N (1987) Parameter orthogonality and approximate conditional
  inference (with discussion). Journal of the Royal Statistical Society Series
  B (Methodological) 49(1):1--39

\bibitem[{Davison and Smith(1990)}]{davison1990models}
Davison AC, Smith RL (1990) Models for exceedances over high thresholds (with
  discussion). Journal of the Royal Statistical Society Series B
  (Methodological) 52(3):393--442

\bibitem[{Efron and Hinkley(1978)}]{efron1978assessing}
Efron B, Hinkley DV (1978) Assessing the accuracy of the maximum likelihood
  estimator: Observed versus expected fisher information. Biometrika
  65(3):457--483

\bibitem[{Gander(1985)}]{gander1985halley}
Gander W (1985) On {H}alley's iteration method. American Mathematical Monthly
  92(2):131--134

\bibitem[{Hills and Smith(1992)}]{hills1992parameterization}
Hills SE, Smith AF (1992) Parameterization issues in {B}ayesian inference.
  Bayesian Statistics 4:227--246

\bibitem[{Hurrell et~al(2003)Hurrell, Kushnir, Ottersen, and
  Visbeck}]{hurrell2003overview}
Hurrell JW, Kushnir Y, Ottersen G, Visbeck M (2003) An overview of the {N}orth
  {A}tlantic oscillation. Geophysical Monograph-American Geophysical Union
  134:1--36

\bibitem[{Northrop and Attalides(2016)}]{northropproper}
Northrop PJ, Attalides N (2016) Posterior propriety in {B}ayesian extreme value
  analyses using reference priors. Statistica Sinica 26(2):721--743

\bibitem[{Northrop and Jonathan(2011)}]{northrop2011threshold}
Northrop PJ, Jonathan P (2011) Threshold modelling of spatially dependent
  non-stationary extremes with application to hurricane-induced wave heights.
  Environmetrics 22(7):799--809

\bibitem[{Pickands(1975)}]{pickands1975statistical}
Pickands J (1975) Statistical inference using extreme order statistics. The
  Annals of Statistics 3(1):119--131

\bibitem[{Robert and Casella(2009)}]{robert2009introducing}
Robert C, Casella G (2009) Introducing Monte Carlo Methods with R. Springer
  Science \& Business Media

\bibitem[{Roberts et~al(2001)Roberts, Rosenthal et~al}]{roberts2001optimal}
Roberts GO, Rosenthal JS, et~al (2001) Optimal scaling for various
  {M}etropolis-{H}astings algorithms. Statistical Science 16(4):351--367

\bibitem[{Smith(1985)}]{smith1985maximum}
Smith RL (1985) Maximum likelihood estimation in a class of nonregular cases.
  Biometrika 72(1):67--90

\bibitem[{Smith(1987{\natexlab{a}})}]{smith1987parameter}
Smith RL (1987{\natexlab{a}}) Discussion of ``{P}arameter orthogonality and
  approximate conditional inference" by {D}.{R}. {C}ox and {N}. {R}eid. Journal
  of the Royal Statistical Society Series B (Methodological) 49(1):21--22

\bibitem[{Smith(1987{\natexlab{b}})}]{smithreport}
Smith RL (1987{\natexlab{b}}) A theoretical comparison of the annual maximum
  and threshold approaches to extreme value analysis. Technical Report 53,
  University of Surrey

\bibitem[{Smith(1989)}]{smith1989extreme}
Smith RL (1989) Extreme value analysis of environmental time series: an
  application to trend detection in ground-level ozone. Statistical Science
  4(4):367--377

\bibitem[{Stephenson(2016)}]{stephenson2016bayesian}
Stephenson A (2016) Bayesian inference for extreme value modelling. Extreme
  Value Modeling and Risk Analysis: Methods and Applications pp 257--280

\bibitem[{Tawn(1987)}]{tawn1987parameter}
Tawn JA (1987) Discussion of ``{P}arameter orthogonality and approximate
  conditional inference" by {D}.{R}. {C}ox and {N}. {R}eid. Journal of the
  Royal Statistical Society Series B (Methodological) 49(1):33--34

\bibitem[{Wadsworth et~al(2010)Wadsworth, Tawn, and
  Jonathan}]{wadsworth2010accounting}
Wadsworth JL, Tawn JA, Jonathan P (2010) Accounting for choice of measurement
  scale in extreme value modeling. The Annals of Applied Statistics
  4(3):1558--1578

\end{thebibliography}

% \begin{thebibliography}{}
% %
% % and use \bibitem to create references. Consult the Instructions
% % for authors for reference list style.
% % Format for Journal Reference
% Author, Article title, Journal, Volume, page numbers (year)
% Author, Book title, page numbers. Publisher, place (year)
% \end{thebibliography}

\end{document}